\documentclass[11pt]{article}
\usepackage[utf8]{inputenc}
\usepackage[utf8]{inputenc}
\usepackage[english]{babel}
\usepackage[T1]{fontenc}
\usepackage{amsmath}
\usepackage{amssymb}

\usepackage{graphicx}

\usepackage{cite}

\usepackage{color} 
\usepackage{graphicx}
\usepackage{authblk}

\usepackage[round]{natbib} 

\usepackage[labelfont=bf,labelsep=period,justification=raggedright]{caption}

\makeatletter
\renewcommand{\@biblabel}[1]{\quad#1.}
\makeatother

\date{}

\title{Imperfect Testing of Individuals for Infectious Diseases: Mathematical Model and Analysis}
\author[1]{Daniel A. M. Villela\\ 
{\tt dvillela@fiocruz.br}\\
Programa de Computação Científica, Fundação Oswaldo Cruz \\ 
Rio de Janeiro, Brazil}

\date{}

\begin{document}

{

\maketitle


\section*{Abstract}

Testing symptomatic individuals for a disease can deliver treatment resources, if tests' results turn positive, which speeds up their treatment and might also decrease individuals' contacts to other ones.  
An imperfect test, however, might incorrectly consider susceptible individuals to be infected (false positives).  In this case, testing reduces the epidemic in the expense of potentially misclassifying individuals.
We present a mathematical model that describes the dynamics of an infectious disease and its testing.
Susceptible individuals turn to ``susceptible but deemed infected'' at rate $\theta$.  
Infected individuals go to a state ``infected and tested positive'' at rate $\alpha$.  Both rates are functions of test's sensitivity and specificity. 
Analysis of the model permits us to derive an expression for $R_0$ and to find the conditions for reaching $R_0<1$, i.e., when the disease--free equilibrium is stable.  
We also present numerical results to cover interesting scenarios such as using different tests and to compare these results.
We find for different sensitivity and specificity values the conditions permitting to get the basic reproduction number $R_0<1$, when originally, i.e., without testing, we would have $R_0>1$. We also find for a given sensitivity and specificity, the critical testing rate for reaching $R_0 <1$.



\section{Introduction}

Treatment for communicable diseases can be quite costly.  In the case of multiresistant tuberculosis infection the treatment per patient hast a cost that is much higher than an infection from non-resistant strain of the pathogen \citep{cohen2009mathematical}.
This kind of treatment, however, relies on testing diagnostics which have
limitations.  They basically might be more accurate at the expense of time taken to have the results.  It is often best to start treatment as soon as the case is diagnosed, but the test results might suffer from the test imperfections.  
A model that indicates the cost of an epidemic as a function of the test 
sensitivity and specificity can help mitigate or even avoid an epidemic.

The test imperfections appear from the probabilities of finding true positives or true negatives.
The first effect comes from the specificity, i.e., the probability that the test indicates a negative diagnosis given that the tested individual does not have the pathogen.
If the specificity is not so high a significant fraction might be considered infected, whereas those individuals are not.  In some cases, starting treatment is the priority over waiting for more accurate results.
Therefore imperfections here might increase the cost.

Also, for real infected individuals, it is important that the test has good sensitivity, i.e., its rate of positive diagnosis in the case of positive tested individuals should be high.  Again, this comes at an expense of rapid diagnosis.
The effect is that individuals that are correctly diagnosed start treatment and reduce both the time they could infect other people and the contacts to susceptible individuals, because of isolation and other measures.


In the case of tuberculosis (TB), for instance, an example is the Xpert (R) MTB/Rif assay test for diagnosis of TB
developed for the GeneXpert platform \citep{who2011rapid}.
It consists of a molecular testing that can be done on demand and closer to individuals in need. 
A meta-analysis of these reports was done by \cite{walusimbi2013meta}. The test results take generally about three hours, whereas a more established protocol (sequence of tests and surveys) would result in treatment starting only after a few days.
\cite{walusimbi2013meta} found pooled specificity and sensitivity estimated at 0.67 (CI: 0.62 -- 0.71) and 0.98 (CI: 0.97--0.99), respectively.

In the case of hepatitis C, \cite{de2014performance} report results for rapid tests which can have high values for both specificity and sensitivity, but can vary depending on the kind of sample (serum, whole blood, saliva).

\cite{eaton2014proportion} estimate the effect of early--stage transmission of HIV, {\it i.e.}, before testing individuals for HIV infection, a stage at which they did not started receiving treatment.

The recent literature on disease modeling and diagnosis is extensive. We point to the review by 
\cite{zwerling2015mathematical} on modeling of TB, investigating diagnosis and therapies. 
In particular, \cite{raimundo2014theoretical} have a an approach on modeling focused on resistant TB. \cite{salje2014importance} discuss the implementation of testing for multiresistant TB in India.
Also, \cite{cox2014benefits} in a recent editorial described the need for simple models for the sake of interpretation of results, especially for people from outside of modeling. In our model, we consider the minimum number of variables to give important insights.



In this paper we present a modeling approach that permits us to obtain $R_0$ as a function of testing specificity and testing sensitivity.  Our model is described as a set of ordinary differential equations, close to a SIR model. We present analytical results from the model and also some numerical simulations that show the cost increase due to testing along with the desired effect of the mitigiation or avoidance of epidemics.

\section{Model}

Our ODE model, leveraged from a classical SIR model, has two components: $S_m$, to describe the number of susceptible individuals that take the test and are deemed incorrectly to be infected, and $I_m$, that describes the number of infected individuals that are test--diagnosed positive and start treatment. 
As usual, we describe the other compartments: $S$, to describe number of suscetibles, $I$ for number of infected individuals and $R$ for recovered ones. 


Individuals that are susceptible get infected at rate $\beta$.  These individuals are called the regular susceptibles.  The model describes the situation in which susceptible individuals take a test and are deemed incorrectly to be infected at rate $\beta_m$.   This class is for the ``susceptible but deemed infected''.  We generally assume that $\beta > \beta_m$, for infected ones might be under treatment, possibly in isolation, therefore these individuals are less likely to transmit the pathogen to another individual.

Individuals leave the population at rate $\mu$, equal for all classes, because we assume a disease that does not cause significantly higher mortality rate.
The rate of new individuals entering the population ({\it e.g.}, typically by birth) 
is also given by $\mu$ to result in a constant population size.  We normalize all variables by the population size.  

Individuals pass from susceptible to a susceptible--deeemed--infected state at rate $\theta$.
This will depend on the rate $r$ at which individuals are tested and the test specificity $\epsilon$. The test specificity is given by the probability of true negatives, i.e. P(test- | uninfected) \citep{newman2003biostatistical}. Hence, $\theta$ is the product of the rate $r$ and P(test+ | uninfected)=$1-\epsilon$, such that $\theta = r (1 - \epsilon)$.

The test sensitivity $\psi$ is given by the rate of true positives, i.e. P(test+ | infected) \citep{newman2003biostatistical}.
Also, infected individuals can be tested and, if the test is positive, start treatment.  Therefore, infected individuals will pass from class $I$ of regular infected individuals to the class of $I_m$ infected under treatment at rate $\alpha= r \psi$.
Finally, we assume that individuals that are under treatment get to recover faster.  Therefore, regular infected individuals recover at a rate $\gamma$, whereas infected--under--treatment individuals recover at rate $\gamma_m$.  Typically, $\gamma_m > \gamma$.

We present the ODE system that describes this model:
\begin{align}
\frac{dS}{dt} & = - \theta S - \beta S(I+ I_m) -\mu S + \mu \nonumber \\ 
\frac{dS_m}{dt} & = \theta S - \beta_m S_m(I+I_m) - \mu S_m \nonumber \\
\frac{dI}{dt} & = \beta S(I+I_m) + \beta_m S_m( I+I_m) - (\alpha +\gamma + \mu) I \label{eq:syst} \\
\frac{dI_m}{dt} & = \alpha I - \gamma_m I_m -\mu I_m \nonumber\\
\frac{dR}{dt} & = \gamma_m I_m + \gamma I - \mu R \nonumber
\end{align}







Table \ref{tab:model} contains a summary of the variables and parameters, and their respective descriptions as well as values used in this work for investigation of numerical scenarios.

\begin{table}[!htbp]
\caption{{\bf Model parameters, variables and respective descriptions.} }
\label{tab:model}
\begin{center}
\begin{tabular}{|c|l|l|}
\hline
Parameters & Description & Value \\ \hline
$\beta$ & transmission rate in state $S$ & .15\\
$\gamma$ & recovery rate in state $I$ & .1\\
$\gamma_m$ & recovery rate in state $I_m$ & .15\\
$\beta_m$ & transmission rate in state $S_m$& .1\\
$\mu $ & mortality/birth rate & 0.003\\ 
$r$ & testing rate & different values \\
$\psi$ & test sensitivity & different values \\
$\epsilon$ & test specificity &  different values\\
$\alpha = r \psi$ & rate at which individuals enter state $I_m$ & different values\\
$\theta = r (1-\epsilon)$ & rate for entering state $S_m$ & different values\\ \hline \hline
Variable & & \\ \hline
$S$ &  number of susceptible individuals & normalized\\
$S_m$ & number of susceptible individuals but deemed infected & normalized\\
$I$ & number of infected individuals & normalized \\
$I_m$ & number of tested--positive infected individuals & normalized\\
$R$ & number of recovered individuals& normalized\\
\hline
\end{tabular}
\end{center}
\end{table}

\subsection*{Cost of infection}

We apply here a model in which treatment cost varies linearly with the number $I$ of infected persons and the number $I_m$, but also the number $S_m$ of susceptible but deemed infected. The cost is a function of the rate $r$ and time $\tau$, given by $C_a(r) = \int_0^{\tau} w_S S_m(r,t) + w_I I(r,t) + w_M I_m(r,t) dt$.
The relative cost $C(\tau)$ is given by the ratio between the cost under rate $r$ and the cost at no treatment, i.e., $r=0$, as follows:
\begin{equation}
\label{eq:costratio}
C(\tau) =  \frac{\int_0^{\tau} w_S S_m(r,t)+ w_I I(r,t)+w_M I_m(r,t) dt}{\int_0^{\tau} w_S S_m(0,t)+ w_I I(0,t)+w_M I_m(0,t) dt},
\end{equation}
where $w_S$, $w_I$, and $w_M$ are weights to each of the variables $S_m$, $I$, and $I_m$, respectively.

\section{Results}

\subsection*{System dynamics}


We start the analysis by considering that there are only susceptible individuals, without any infectious individuals.  In this case, we have clearly $I(t)=0$, $I_m(t)=0$ and the left hand side of equations in \ref{eq:syst} become zero.
The solution is given by
\begin{equation*}
(S_f,S_{m,f},I_f,I_{m,f},R_f) = ( \frac{\mu}{\mu + \theta},  \frac{\theta}{\mu + \theta}, 0, 0, 0).
\end{equation*}

As noted above,
he system has an equilibrium in which $S_m = \frac{\theta}{\mu + \theta}$.
Therefore, a fraction of the population demands treatment resources even when there are no infected individuals.


We follow from this initial result to derive a matrix that describes the ``flows'' into infectious states and out from infectious states.
Note that the state $S_m$ is not really an infectious state, since the individual is still susceptible.
We apply the method of analyzing the Next-Generation Matrix (NGM, described by \cite{diekmann2009construction}) $\mathbf{K}$ and find:

\begin{equation*}
\mathbf{K}=
\left(\begin{array}{rr}
\frac{\frac{\beta \mu}{\mu + \theta} + \frac{\beta_{m} \theta}{\mu + \theta}}{\alpha + \gamma + \mu} + \frac{\alpha {\left(\frac{\beta \mu}{\mu + \theta} + \frac{\beta_{m} \theta}{\mu + \theta}\right)}}{{\left(\alpha + \gamma + \mu\right)} {\left(\gamma_{m} + \mu\right)}} & \frac{\frac{\beta \mu}{\mu + \theta} + \frac{\beta_{m} \theta}{\mu + \theta}}{\gamma_{m} + \mu} \\
0 & 0
\end{array}\right).
\end{equation*}


The dominant eigenvalue of the matrix $\mathbf{K}$ given by the spectral radius of matrix $\mathbf{K}$ results in the basic reproduction number $R_0$, the number of individuals that get infected upon a single infected individual:
%

\begin{equation}
R_0 = \frac{\beta \mu + \beta_m \theta}{(\theta + \mu)(\gamma+\mu+\alpha)}(1+\frac{\alpha}{\gamma_m+\mu}).
\end{equation}


\subsection*{Endemic state}
 
Analysis of the system dynamics permits us to show that it reaches
equilibrium in en endemic state, as expected depending on the threshold given by $R_0$.  The endemic state is given by:
\begin{eqnarray*}
S^* & = & \frac{\mu}{\theta+\mu+\beta I^* (1+\alpha/(\gamma_m+\mu))}\\
S_m^* & = & \left(\frac{\theta}{\beta_m I^* (1+\alpha/(\gamma_m+\mu)) + \mu}\right) \left(\frac{\mu}{\theta + \mu + \beta I^* (1 + \alpha/(\gamma_m + \mu))}\right)\\
I_m^* & = & \frac{\alpha I^*}{\gamma_m+\mu}\\
R^*  & = & \frac{\gamma_m \frac{\alpha}{\gamma_m+\mu} + \gamma}{\mu}I^{*}
\end{eqnarray*}

Using the equations above we can find an equation containing only the number $I^*$ of infected in the endemic state.  
Solving the equation for $I^*$ requires finding the solutions to the polynomial
\begin{equation}
\label{eq:pol}
p_I(I) = \frac{(\theta+\mu) \mu}{\beta \beta_m (1 + \alpha/(\gamma_m+\mu))^2} (R_0-1) + I  \left(\frac{\mu}{\gamma+\alpha+\mu} - \frac{(\theta+\mu+\beta \mu/\beta_m)}{\beta (1+\alpha/(\gamma_m+\mu))}\right) - I^2
\end{equation}
Clearly, $R_0>1$ is a sufficient condition for a biologically feasible solution (a nonnegative real root).
In this case, 
\begin{equation}
I^{*} =  \frac{\sqrt{b^2 + 4 u (R_0 -1)}}{2} - \frac{b}{2} 
\end{equation}
where $b= \frac{\mu}{\gamma+\alpha+\mu} - \frac{(\theta+\mu+\beta \mu/\beta_m)}{\beta (1+\alpha/(\gamma_m+\mu))}$
and 
$u= \frac{(\theta+\mu) \mu}{\beta \beta_m (1 + \alpha/(\gamma_m+\mu))^2}$.

\subsection*{Conditions involving the testing rate}


The rates $\theta$ and $\alpha$ at which individuals are, respectively, deemed infected but still susceptible and 
diagnosed infected both depend on a treatment rate $r$.  These rates will then depend on the testing rate and the test specificity and sensitivity.
By applying the rates $\alpha=r \psi$ and $\theta = r (1-\epsilon)$ depending on the rate $r$, and the specificity $\epsilon$ and sensitivity $\psi$, we rewrite $R_0$:
%
\begin{equation*}
\label{eq:r0simple}
R_0 = \frac{(\beta_m r (1-\epsilon) + \beta \mu)(r \psi + \gamma_m + \mu)}{(r (1-\epsilon) +\mu)(r \psi + \gamma + \mu)(\gamma_m+\mu)}.
\end{equation*}

It is interesting to note that if $r \rightarrow \infty$, $R_0 \rightarrow \frac{\beta_m (1-\epsilon)\psi}{(\gamma_m + mu) \psi (1-\epsilon)}= \frac{\beta_m}{\gamma_m+\mu}$.
This expression is the equal to the $R_0$ obtained for a SIR system at an infection rate $\beta_m$ and recovery rate $\gamma_m$.
When $r=0$, we have a regular SIR model formulated with parameters $\beta$, $\gamma$ and $\mu$, such that $R_0=\frac{\beta}{\gamma+\mu}$. Therefore, these are the two extreme values for $R_0$ when varying the testing rate $r$.




\subsection*{ Conditions for $\mathbf{R_0 < 1}$}

We consider now conditions for the disease not to go to the endemic state, {\it i.e.}, for which $R_0<1$.
This condition would require that a polinomial $p(r)<0$, where
\begin{eqnarray*}
p(r) & = & (1-\epsilon) \psi (\beta_m - \gamma_m - \mu) r^2 + \nonumber\\
 & & (1-\epsilon) (\gamma_m + \mu) (\beta_m - \gamma - \mu) + \mu \psi (\beta - \gamma_m -\mu) r + \nonumber\\
& & \mu (\gamma_m + \mu)(\beta - \gamma - \mu). \nonumber
\end{eqnarray*}

The typical situation for analyzing $p(r) < 0$ is that an epidemic occurs if no treatment is realized, i.e. in this case $R_0>1$, when $r=0$.  Therefore, we would expect the normal SIR threshold condition $\beta > \gamma + \mu$.
Let us first assume that $\beta_m < \gamma_m + \mu$, which is the condition for the polynomial to be concave down.  As shown when $r \rightarrow \infty$, this condition permits an $R_0 <1$, which suggests that under treatment it is possible to bring the system to $R_0=1$ for a finite $r$. 
Indeed, $p(r)$ has two real roots, a positive root $r_c$ and a negative one.
Therefore we would need $r > r_c$ in order to have $p(r)<0$, and consequently $R_0<1$.

If we consider instead an asymptotic condition $\beta_m/(\gamma_m+\mu) >1$, given the reasonable assumptions $\gamma_m > \gamma$ and $\beta > \beta_m$, would guarantee $\beta > \gamma + \mu$, and, by consequence,
\begin{align*}
\beta > \beta_m > \gamma_m + \mu > \gamma + \mu
\end{align*}

In this case, either there are no purely real roots or none of the roots are positive.
Hence, $p(r)>0$, $R_0 >1$.
This essentially means that increasing the testing rate permits to decrease $R_0$, but not to avoid an epidemic.

\begin{figure}[!htb]
\begin{center}
\resizebox*{0.75\textwidth}{!}{\includegraphics{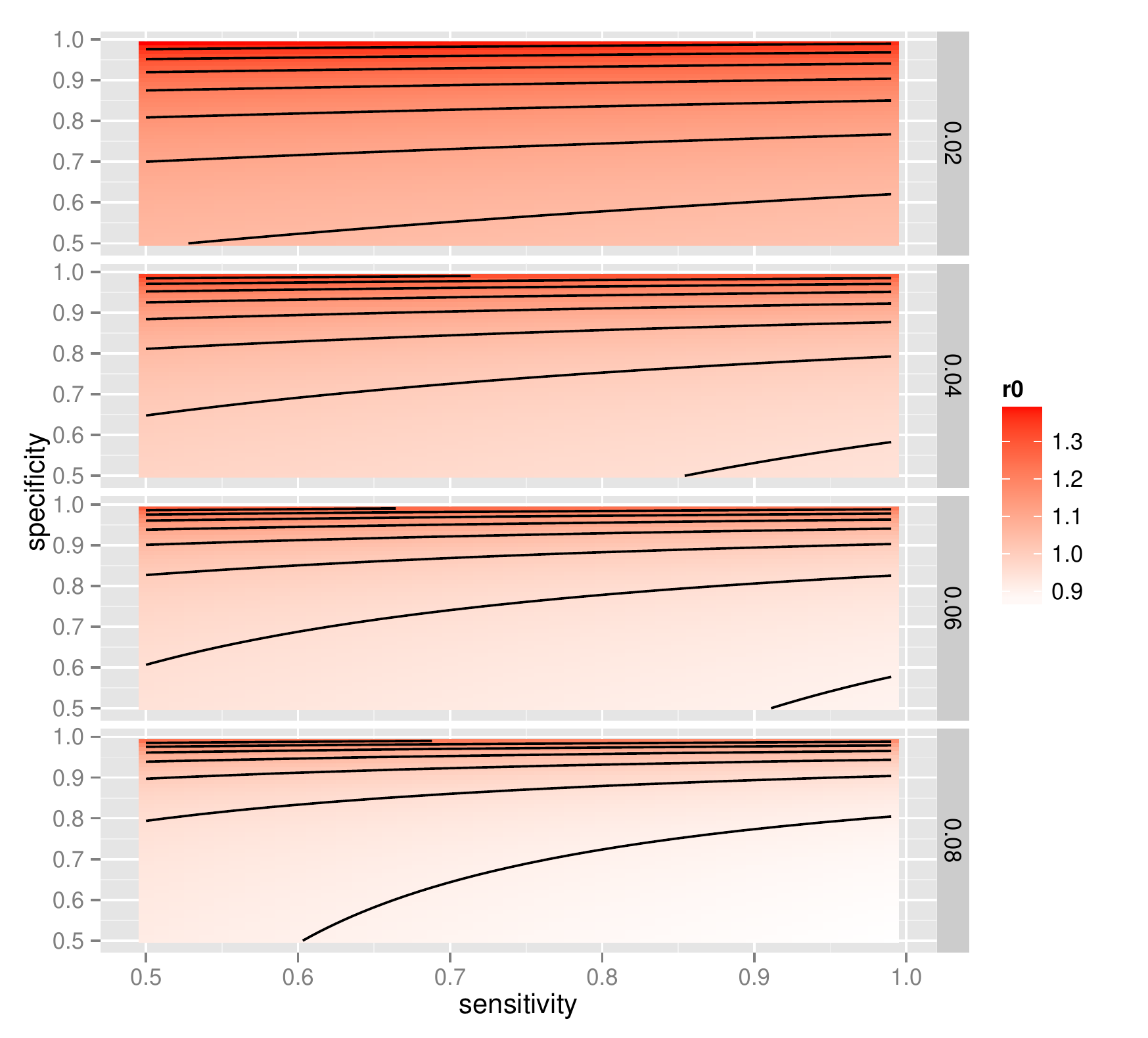}}%
\end{center}
\caption{{\bf Countour plot under different testing rates.} We show here testing rates from top to bottom: $r=0.02$, $r=0.04$, $r=0.06$, $r=0.08$. The intensity of the color indicates the $R_0$: colors closest to white are the smallest $R_0$ values, whereas colors close to red indicate high values for $R_0$.} \label{hsh.fig}
\end{figure}

\subsection*{Different scenarios: numerical simulations}

Here we consider the model as given by Table \ref{tab:model}. Values for parameters $\beta$, $\beta_m$, $\gamma$, $\gamma_m$, $\mu$ were chosen such that we have $R_0$ on a domain that permits situations from epidemic to controlled disease. 
We should note that under no treatment, we have an $R_0$ condition: $R_0 = \frac{\beta}{\gamma+\mu}= 1.456$.
When we have the limiting condition, $r \rightarrow \infty$, we find $R_0 = \frac{\beta_m}{\gamma_m+\mu} = 0.654$.
Therefore these two values are the extreme points that we can find by considering different scenarios for $\theta$ and $\alpha$ (or alternatively different values for $r$, $\epsilon$, and $\psi$).


We first consider varying the sensitivity and specificity for different testing rates given by $r$.
Figure \ref{hsh.fig} shows us from top to bottom the different countour plots for values of the basic reproduction number $R_0$ when varying the testing rate $r$ from $r=0.02$ to $r=0.08$. The intensity of the color permits us to observe the $R_0$ values, in which a lighter-intensity color depicts smaller values for $R_0$.  As expected if we increase the testing rate $r$, the plots become lighter, meaning that epidemics are less likely under higher testing rates.

\begin{figure}[!htb]
\begin{center}
\resizebox*{0.75\textwidth}{!}{\includegraphics{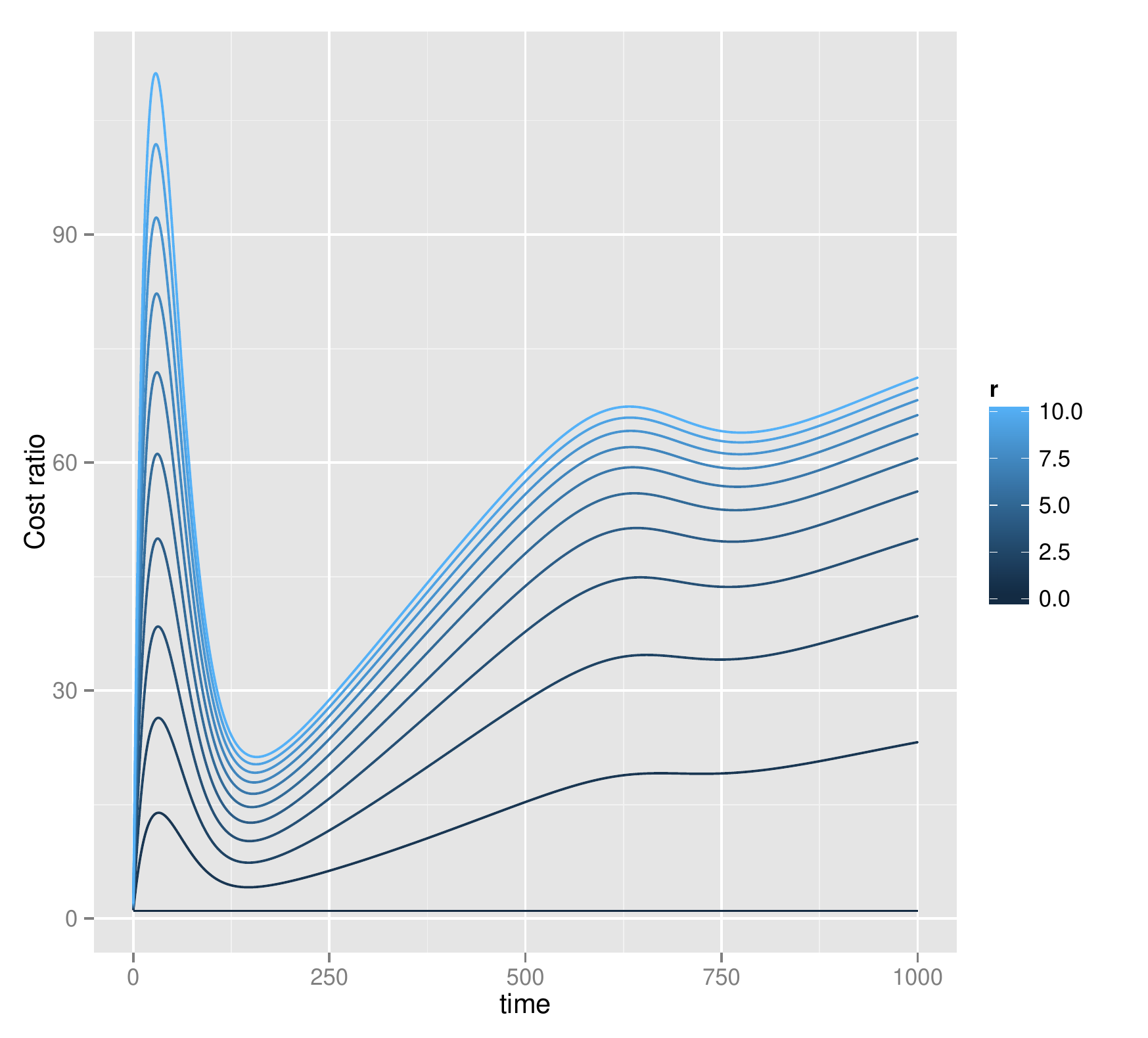}}%
\end{center}
\caption{{\bf Cost ratio.} Each of the curves depict the cost ratio obtained over time for different testing rates $r$.  Curves vary from $r=0$ (dark blue) to $r=10$ (light blue). } \label{fig:costratio}
\end{figure}


We also consider the cost ratio given by Eq. \ref{eq:costratio} and varying the testing rate $r$.  
Here, we consider the sensitivity equal to 0.98 and the specificity at 0.67, values found for a TB test by pooled results \citep{walusimbi2013meta}, even though the model might be simple for analyzing complex scenarios that TB presents.
Figure \ref{fig:costratio} shows multiple curves, in which we observe for higher testing rates (lighter colors) the cost ratio increasing to higher peaks.  This is expected because more individuals are under treatment.

Finally, we show for the same testing rates as in Figure \ref{fig:costratio}, how the epidemic develops (or not). 
We observe in Figure \ref{fig:incidence} that the incidence effectively decreases as the testing rate increases (lighter colors), which is the opposite effect of the cost ratio.  Therefore, the ideal point should be a testing rate $r$ that enables a basic reproduction number $R_0 <1$, but probably close to the threshold such that the cost ratio is not prohibitive.

\begin{figure}[!htb]
\begin{center}
\resizebox*{0.75\textwidth}{!}{\includegraphics{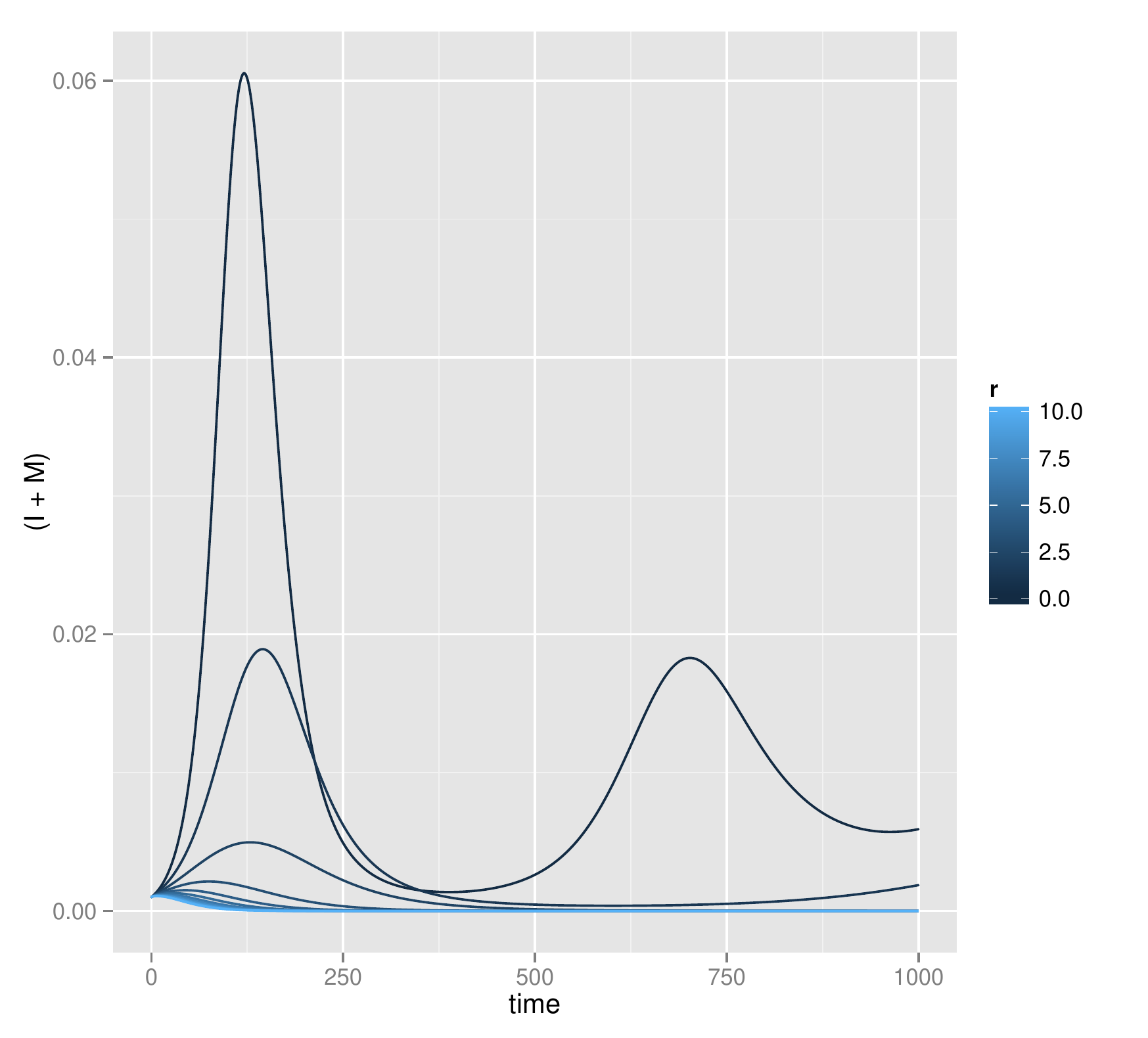}}%
\end{center}
\caption{{\bf Incidence at different testing rates.} Each of the curves show the number of infected persons under different testing rates varying from $r=0$ (dark blue) to $r=10$ (light blue). The number of infected persons are found by summing variables $I$ and $I_m$. } \label{fig:incidence}
\end{figure}



\section{Discussion}

In this paper we establish a mathematical modeling that permits us to analyze the effects of imperfect testing on the dynamics of diseases.
The modeling considers specificity and sensitivity of the testing. 
Therefore, given these two parameters we can find $R_0$, which tells us if the disease--free state is stable.  In this case, even if few disease cases appear, we would not expect an epidemic to occur.
 Alternatively, we can find a testing rate that can bring the disease to a controled level for a specific test, for which specificity and sensitivity has been estimated.

The testing rate, however, also raises the total treatment costs given a time interval.
Hence, as expected, it is not viable to raise the testing rate indefinitely, for cost--effectiveness, especially when the per--capita treatment is already costly.


We believe our model is particularly helpful to give insights about regions (controlled disease/epidemic, cost prohibitive/viable costs).  We expect to extend the work to other similar models, in particular a SEIR model.  
This would be the case for diseases such as tuberculosis, and especially multi--resistant tuberculosis, which requires expensive treatment.


We considered estimated rates for sensitivity and specificity typical of tests, such as a tuberculosis test GenExpert \citep{walusimbi2013meta}, to evaluate possible scenarios.  We also intend to study data from public health systems that show the impact of introducing treatment and the cost associated with the treatment.

The imperfect testing might bring individuals to states, for instance a false--positive infected, in which they might be diagnosed after a slow test and should return to susceptible state.  We intend to treat this case in future work.



We consider the expected cases in which $\beta > \beta_m$, a smaller force of infection under treatment, and $\gamma_m > \gamma$, a speedy recover rate under treatment. Those are reasonable assumptions in most cases.  Different cases, however, can be considered for more complex scenarios.  In the case of sexual--contact diseases, maybe some individuals might incur on reckless behavior, which would instead not increase the treatment, but increase the force of infection, as discussed by \cite{wilson2008relation}. We intend to pursue investigating such ``anomalies'' in the framework of our model.




\bibliographystyle{abbrvnat}
\bibliography{treat} 

\end{document}